\documentclass[prb,twocolumn,groupedaddress,shownopacs,amssymb]{revtex4-2}
\usepackage{graphicx,psfrag,amsmath,amssymb}

\usepackage[usenames, dvipsnames]{color}

\usepackage{hyperref} 
\hypersetup{
    colorlinks=true,
    linkcolor=Blue,
    citecolor=Blue,
    filecolor=Blue,      
    urlcolor=Blue,
    pdftitle={Coherence Length},
    }

\begin{document}

\title{Coherence length and quantum geometry in a dilute flat-band superconductor}

\author{M. Iskin}
\affiliation{
Department of Physics, Ko\c{c} University, Rumelifeneri Yolu, 
34450 Sar\i yer, Istanbul, T\"urkiye
}

\date{\today}

\begin{abstract}
To explore the influence of quantum-geometric effects on the Ginzburg-Landau 
coherence length in a dilute flat-band superconductor, we adopt a BCS-BEC 
crossover approach to the multiband pyrochlore-Hubbard model near the 
critical temperature for superconductivity. Our self-consistent formulation 
for this three-dimensional lattice benchmarks very well against the so-called 
zero-temperature coherence length, demonstrating the monotonic decay of 
the coherence length to zero as the interaction strength increases. 
Additionally, we show that the effective mass of the many-body bound states 
(i.e., Cooper pairs) is nearly identical to that of the lowest-lying 
two-body bound states in the dilute flat-band limit.

\end{abstract}
\maketitle

\section{Introduction}
\label{sec:intro}

Recent theoretical studies have uncovered a surprising link between 
the quantum geometry of Bloch states and superconductivity in multiband 
systems, significantly advancing our understanding of the Cooper-pairing 
mechanism in flat-band systems~\cite{torma22, torma23}. 
Under certain restrictive conditions, such as time-reversal symmetry 
and uniform pairing, it has been revealed that the quantum-metric tensor 
plays a critical role in determining key observables. 
These include the superfluid weight, superfluid density, 
critical transition temperature, low-energy collective Goldstone and 
Leggett modes, Ginzburg-Landau (GL) coherence length, London penetration 
depth, GL parameter, and upper critical magnetic field
~\cite{torma22, torma23, huhtinen22, herzog22, iskin23, iskin24, daido23, 
chen23, thumin23, verma24, jiang24}. 
The common factor among these observables is the effective mass of 
the superfluid carriers~\cite{iskin18b, iskin18c}. 
Despite the infinite band mass of particles in the underlying flat band, 
even an infinitesimal interaction gives Cooper pairs a finite effective 
mass through virtual interband transitions mediated by the quantum metric. 
This mechanism facilitates the emergence of superconductivity in a flat 
band in the presence of other bands~\cite{torma18}. Furthermore, 
the geometric origin is most evident in the effective mass of the 
lowest-lying two-body bound states, as demonstrated by exact
calculations~\cite{iskin22}.

In multiband superconductors, geometric effects, while inherent, are 
generally overshadowed by conventional effects in the BCS 
limit~\cite{torma22, torma23}. 
However, as the bandwidth of a Bloch band narrows, conventional effects 
diminish due to their dependence on the derivative of the Bloch bands. 
In contrast, geometric effects, which are determined by the derivative 
of the Bloch states, can become dominant under these conditions. 
Inspired by recent technological advances in creating two-dimensional 
materials with nearly-flat electronic bands~\cite{balents20}, 
there is a growing surge of interest in understanding geometric 
effects in multiband superconductors~\cite{tian23, wakefiel23, huang23}.
Among them, there has been a perplexing zero-temperature GL study regarding 
the size of Cooper pairs in a flat-band superconductor, suggesting they 
are limited by a fundamental length scale determined by the quantum 
geometry of the flat band~\cite{hu23}. However, subsequent work using
the Bogoliubov-de Gennes approach for various lattice models in one 
and two dimensions has demonstrated that characteristic correlation lengths 
can be smaller than one lattice spacing without being constrained by the quantum 
metric~\cite{thumin24}.

To address these discrepancies~\cite{hu23, thumin24}, we investigate 
the quantum-geometric effects on the GL coherence length in a dilute 
flat-band superconductor using a BCS-BEC crossover approach to the 
multiband pyrochlore-Hubbard model near the critical temperature $T_c$ 
for pairing. Our self-consistent results are in good agreement with the 
zero-temperature coherence length, demonstrating that GL coherence length 
decreases to zero monotonously as the interaction strength increases. 
In addition, we find that the effective mass of Cooper pairs aligns 
closely with that of the lowest-lying two-body bound states in the 
dilute regime. The pyrochlore lattice, akin to well-studied two-dimensional 
models like the Mielke-checkerboard and kagome lattices known for 
uniform pairing, provides an ideal framework for exploring three-dimensional 
flat-band superconductivity~\cite{iskin24}. 
Recent demonstrations of flat bands and superconductivity in materials 
like pyrochlore metal CaNi$_2$~\cite{wakefiel23} and pyrochlore 
superconductor CeRu$_2$~\cite{huang23} underscore the relevance of this model.

The rest of of the paper is organized as follows. In Sec.~\ref{sec:tdgl},
we introduce the multiband Hubbard Hamiltonian in reciprocal space, 
review the effective-action formalism near $T_c$, and derive the 
expansion coefficients for the time-dependent GL (TDGL) theory. 
In Sec.~\ref{sec:sce}, we discuss self-consistency relations that are 
used to study dilute flat-band superconductivity in the pyrochlore lattice. 
In Sec.~\ref{sec:T0}, we introduce the zero-temperature coherence length. 
In Sec.~\ref{sec:num}, we present the numerical results and analyze 
them in various limits. 
In Sec.~\ref{sec:disc}, we comment on the main source of discrepancy. 
The paper ends with a summary and outlook in Sec.~\ref{sec:conc}, 
and the breakdown of zero-temperature coherence length is discussed 
in the Appendix.

\section{TDGL Theory}
\label{sec:tdgl}

Having the pyrochlore-Hubbard model that we recently studied at zero 
temperature ($T = 0$) in mind~\cite{iskin24}, we begin by introducing 
the multiband Hubbard model with an onsite attractive interaction and 
generic hopping terms. We then review the derivation of the effective 
action near the critical temperature $T_c$ for superconductivity 
and apply the resultant TDGL theory to the pyrochlore lattice, 
which exhibits time-reversal symmetry and uniform pairing.

\subsection{Multiband Hubbard Hamiltonian}
\label{sec:Ham}

Just as the kagome lattice is a line graph of the honeycomb lattice and 
the Mielke checkerboard lattice is a line graph of the square lattice, 
both of which feature a flat tight-binding band in two dimensions, 
the pyrochlore lattice is a line graph of the diamond lattice and it 
features two degenerate flat bands in three dimensions. Its crystal 
structure consists of a face-centered-cubic Bravais lattice with a 
four-point basis, leading to a truncated-octahedron shaped Brillouin 
Zone (BZ)~\cite{iskin24}. This is in such a way that the total number 
of lattice sites is $N = 4 N_c$, where
$
N_c = \sum_{\mathbf{k} \in \mathrm{BZ}} 1
$ 
counts the number of unit cells in real space, with 
$\mathbf{k} = (k_x, k_y, k_z)$ denoting the crystal momentum in units 
of $\hbar \to 1$.

Upon Fourier transformation from the site representation to the reciprocal 
space, the multiband Hubbard model becomes
\begin{align}
\mathcal{H} &= \sum_{S S' \mathbf{k} \sigma} \psi_{S \mathbf{k} \sigma}^\dagger 
(h_{SS'}^{\mathbf{k} \sigma} - \mu_\sigma \delta_{SS'}) \psi_{S' \mathbf{k} \sigma}
\nonumber \\
&- \frac{U}{N_c} \sum_{S \mathbf{k} \mathbf{k'} \mathbf{q}} 
\psi_{S \mathbf{k} \uparrow }^\dagger 
\psi_{S, -\mathbf{k}+\mathbf{q}, \downarrow}^\dagger
\psi_{S, -\mathbf{k'}+\mathbf{q}, \downarrow}
\psi_{S \mathbf{k'} \uparrow},
\label{eqn:ham}
\end{align}
where $\psi_{S \mathbf{k} \sigma}^\dagger$ creates a spin 
$
\sigma=\{\uparrow,\downarrow\}
$ 
particle on the sublattice $S = \{A, B, C, D\}$ and $\delta_{ij}$ is a
Kronecker delta. Here, $U \ge 0$ is the strength of the attractive onsite 
interaction between $\uparrow$ and $\downarrow$ particles, and $\mu_\sigma$ 
is the chemical potential determining their average numbers in the ensemble. 
In the specific case of the pyrochlore lattice, which is our primary 
interest in this paper, the matrix elements of the Bloch Hamiltonian 
$
\mathbf{h}_{\mathbf{k} \sigma}
$ 
can be written as
$
h_{SS}^{\mathbf{k} \sigma} = 0,
$
$
h_{AB}^{\mathbf{k} \sigma} = -2\bar{t} \cos\big(\frac{k_y+k_z}{4}a\big),
$
$
h_{AC}^{\mathbf{k} \sigma} = -2\bar{t} \cos\big(\frac{k_x+k_z}{4}a\big),
$
$
h_{AD}^{\mathbf{k} \sigma} = -2\bar{t} \cos\big(\frac{k_x+k_y}{4}a\big),
$
$
h_{BC}^{\mathbf{k} \sigma} = -2\bar{t} \cos\big(\frac{k_x-k_y}{4}a\big),
$
$
h_{BD}^{\mathbf{k} \sigma} = -2\bar{t} \cos\big(\frac{k_x-k_z}{4}a\big)
$
and
$
h_{CD}^{\mathbf{k} \sigma} = -2\bar{t} \cos\big(\frac{k_y-k_z}{4}a\big),
$
where $\bar{t}$ is the tight-binding hopping parameter between the 
nearest-neighbor sites and $a$ is the side-length of the conventional 
simple-cubic cell~\cite{iskin24}. Thus, the Bloch bands are determined by 
$
\sum_{S'} h_{SS'}^{\mathbf{k} \sigma} n_{S' \mathbf{k} \sigma} 
= \varepsilon_{n\mathbf{k} \sigma} n_{S \mathbf{k} \sigma},
$
leading to
$
\varepsilon_{1\mathbf{k}\sigma} = -2\bar{t}(1 + \sqrt{1 + \alpha_\mathbf{k}})
$
and
$
\varepsilon_{2\mathbf{k}\sigma} = -2\bar{t}(1 - \sqrt{1 + \alpha_\mathbf{k}})
$
for the dispersive bands, where 
$
\alpha_\mathbf{k} = \cos(k_x a/2) \cos(k_y a/2) + 
\cos(k_y a/2) \cos(k_z a/2) + \cos(k_x a/2) \cos(k_z a/2),
$
and
$
\varepsilon_{3\mathbf{k}\sigma} = \varepsilon_{4\mathbf{k}\sigma} = 2\bar{t}
$
for the degenerate flat bands. Note that $\varepsilon_{2\mathbf{k}\sigma}$ 
touches the flat bands at $\mathbf{k = 0}$. 
To ensure that these flat bands appear at the 
bottom of the Bloch spectrum, we set $\bar{t} \to -t$ and choose $t > 0$ as 
the unit of energy. This allows us to construct a relatively simple BCS-BEC 
crossover formalism for a dilute flat-band superconductor near $T_c$, and 
study some of its properties as functions of $U$. Here, 
$
n_{S \mathbf{k} \sigma} = \langle S | n_{\mathbf{k} \sigma} \rangle
$
is the sublattice projection of the Bloch state $| n_{\mathbf{k} \sigma} \rangle$, 
which plays an important role throughout this paper.

\subsection{Effective action near $T \le T_c$}
\label{sec:ea}

In order to construct a microscopic TDGL theory near $T_c$~\cite{sademelo93}, 
we first extract the effective free-energy density of the system per lattice 
site as a function of the pairing order parameter from the effective action, 
i.e., from
$
\frac{T}{N} (\mathcal{S}_2 + \mathcal{S}_4),
$
where $T$ is the temperature in units of the Boltzmann constant 
$k_\mathrm{B} \to 1$, and $\mathcal{S}_2$ ($\mathcal{S}_4$) is the quadratic 
(quartic) contribution to the action~\cite{iskin23}. The saddle-point action 
$\mathcal{S}_0$ is not needed here because it does not depend on the order 
parameter at $T_c$, and $\mathcal{S}_1$ and $\mathcal{S}_3$ vanish at any $T$ 
due to the saddle-point condition.

Using the Grassmann functional-integral formalism, 
one can show that
$
\mathcal{S}_2 = \frac{N_c}{T} \sum_{S S' q} \Lambda_S^*(q) \Gamma^{-1}_{SS'}(q) \Lambda_{S'} (q)
$
~\cite{iskin23}, where
$
q \equiv (\mathbf{q}, \mathrm{i}\nu_\ell)
$ 
is a collective index with $\nu_\ell = 2\ell \pi T$ denoting the bosonic 
Matsubara frequency, $\Lambda_S(q)$ plays the role of the pairing order 
parameter for sublattice $S$, and
\begin{align}
\Gamma_{SS'}^{-1}(q) = \frac{\delta_{SS'}}{U} &+ \frac{1}{2 N_c} \sum_{nm\mathbf{k}} 
\frac{\mathcal{X}_{n \mathbf{k} \uparrow} + \mathcal{X}_{m, -\mathbf{k}+\mathbf{q}, \downarrow}}
{\mathrm{i}\nu_\ell - \xi_{n \mathbf{k} \uparrow} -  \xi_{m,-\mathbf{k}+\mathbf{q}, \downarrow}}
\nonumber \\ 
& \times n_{S \mathbf{k} \uparrow} n_{S' \mathbf{k} \uparrow}^* 
m_{S', -\mathbf{k}+\mathbf{q}, \downarrow}^* m_{S, -\mathbf{k}+\mathbf{q}, \downarrow},
\label{eqn:GammaSS}
\end{align}
is the matrix element of the inverse pair-fluctuation propagator
$
\boldsymbol{\Gamma}^{-1}(q).
$
Here,
$
\mathcal{X}_{n\mathbf{k}\sigma} = \tanh\big(\frac{\xi_{n\mathbf{k}\sigma}}{2T}\big) 
$
is a thermal factor and 
$
\xi_{n \mathbf{k} \sigma} = \varepsilon_{n \mathbf{k} \sigma} - \mu_\sigma.
$
This propagator suggests that the generalized Thouless condition 
$
\det \boldsymbol{\Gamma}^{-1}(\mathbf{q}, 0) = 0
$
determines $T_c$ of a multiband Hubbard model for any center-of-mass 
momentum $\mathbf{q}$. For instance, 
$
T_c = \max\{ T_{c_1}, T_{c_2}, T_{c_3}, T_{c_4}\}
$
in the case of the pyrochlore lattice, where $T_{c_j}$ is determined by 
setting the $j$th eigenvalue of $\boldsymbol{\Gamma}^{-1}(\mathbf{q}, 0)$ 
to $0$. In our numerical calculations, we observed that 
$
T_{c_1} > T_{c_2} = T_{c_3} = T_{c_4}
$ 
for any given set of $(\mu_\uparrow = \mu_\downarrow, U)$ in the 
$\mathbf{q} \to \mathbf{0}$ limit. Furthermore, we observed that 
the associated eigenvector of $\boldsymbol{\Gamma}^{-1}(\mathbf{q}, 0)$ 
that corresponds to the $T_{c_1}$ solution is uniform in a unit cell, 
i.e.,
$
[\Lambda_A(\mathbf{q}), \Lambda_B(\mathbf{q}), 
\Lambda_C(\mathbf{q}), \Lambda_D(\mathbf{q})] = 
\Lambda_0(\mathbf{q}) [1,1,1,1]
$
in the $\mathbf{q} \to \mathbf{0}$ limit.
It is pleasing to note that the latter observation aligns perfectly 
with our previous finding that the uniform-pairing condition is 
satisfied exactly for the lowest-lying two-body bound states of the
pyrochlore lattice when $\mathbf{q \to 0}$~\cite{iskin24}.

Motivated by these numerical insights and to facilitate further 
analytical progress, we next adopt the following assumptions that 
are satisfied by the pyrochlore lattice. 
($i$) The lattice manifests time-reversal symmetry, leading to
$
n_{S, -\mathbf{k}, \downarrow}^* = n_{S \mathbf{k} \uparrow} 
= n_{S \mathbf{k}} 
$
and 
$
\xi_{n, -\mathbf{k}, \downarrow} = \xi_{n \mathbf{k} \uparrow} 
= \xi_{n \mathbf{k}}.
$
($ii$) The low-$q$ order parameters are uniform in a unit cell, 
leading to $\Lambda_0(q) = \Lambda_S(q)$ for all sublattices. 
Under these assumptions, the quadratic action can be written 
as
$
\mathcal{S}_2 = \frac{N}{T} \sum_{q} \Gamma^{-1}_0(q) |\Lambda_0 (q)|^2
$
~\cite{iskin23}, where
\begin{align}
\Gamma^{-1}_0(q) &= \frac{1}{U} + \frac{1}{2 N} \sum_{nm\mathbf{k}}
\frac{
\mathcal{X}_{n \mathbf{k}} + \mathcal{X}_{m, \mathbf{k-q}} }
{\mathrm{i}\nu_\ell - \xi_{n\mathbf{k}} - \xi_{m, \mathbf{k-q}}}
|\langle n_\mathbf{k} | m_\mathbf{k-q} \rangle|^2.
\label{eqn:Gamma0}
\end{align}
Thus, the Thouless condition reduces to
$
\Gamma_0^{-1}(\mathbf{0}, 0)= 0
$
for stationary BCS-type pairing, determining $T_c$ of a uniformly-paired 
multiband Hubbard model in the presence of time-reversal symmetry. 
Similarly, by making use of the Grassmann functional-integral formalism, 
and under the same assumptions discussed above, one can approximate that
$
\mathcal{S}_4 = \frac{N b_0}{2 T}
\sum_{q_1 q_2 q_3}
\Lambda_0^*(q_1+q_2+q_3)\Lambda_0(q_1) \Lambda_0^*(-q_2) \Lambda_0(q_3)
$
~\cite{iskin23}. 
Here, a positive $b_0$ coefficient not only guarantees the energetic 
stability of the TDGL theory, but it also characterizes the repulsive
interaction between Cooper pairs as discussed next.

\subsection{Coefficients for the TDGL expansion}
\label{sec:coeff}

When $T \lesssim T_c$, the coefficients for the microscopic TDGL theory 
can be determined through the expansion of the inverse propagator
$
\Gamma^{-1}_0(\mathbf{q}, \mathrm{i}\nu_\ell \to \omega + \mathrm{i}0^+) = 
- a_0 \epsilon (T) + \frac{1}{2} \sum_{ij} c_{ij} q_i q_j + d \omega + \cdots
$
in the low-$(\mathbf{q}, \omega)$ regime, where 
$
\epsilon (T) = (T_c - T)/T_c
$
changes sign across $T_c$ in accordance with the Landau theory of 
second-order phase transitions~\cite{iskin23, sademelo93}. 
This expansion leads to
\begin{align}
a_0 &= \frac{1}{N} \sum_{n \mathbf{k}} \bigg[
\frac{\mathcal{Y}_{n \mathbf{k}}}{4T_c} 
+ \frac{\partial \mu}{\partial T} 
\bigg( \frac{\mathcal{Y}_{n \mathbf{k}}}{4\xi_{n \mathbf{k}}} 
- \frac{T_c \mathcal{X}_{n \mathbf{k}}}{2 \xi_{n \mathbf{k}}^2}  \bigg)
\bigg],
\label{eqn:a0}
\\
b_0 &= \frac{1}{N} \sum_{n\mathbf{k}} \left(
\frac{\mathcal{X}_{n \mathbf{k}}}{4\xi_{n\mathbf{k}}^3}
- \frac{\mathcal{Y}_{n\mathbf{k}}}{8T_c\xi_{n\mathbf{k}}^2}
\right),
\label{eqn:b0}
\\
c_{ij}^\mathrm{intra} &= \frac{1}{N} \sum_{n \mathbf{k}} \left( 
\frac{\mathcal{X}_{n\mathbf{k}}}{4\xi_{n\mathbf{k}}^3} 
- \frac{\mathcal{Y}_{n\mathbf{k}}}{8T_c\xi_{n\mathbf{k}}^2} 
\right) \dot{\xi}_{n\mathbf{k}}^i \dot{\xi}_{n\mathbf{k}}^j,
\label{eqn:cintra}
\\
c_{ij}^\mathrm{inter} &= \frac{1}{N} \sum_{n \mathbf{k}}
\frac{\mathcal{X}_{n\mathbf{k}}}{2\xi_{n\mathbf{k}}} g^{n\mathbf{k}}_{ij}
-  \frac{1}{2N} \sum_{n, m \ne n, \mathbf{k}} \frac{\mathcal{X}_{n\mathbf{k}} + \mathcal{X}_{m\mathbf{k}}}
{\xi_{n\mathbf{k}}+\xi_{m\mathbf{k}}} g^{nm\mathbf{k}}_{ij},
\label{eqn:cinter}
\\ 
d &=  \frac{1}{N} \sum_{n \mathbf{k}} \frac{\mathcal{X}_{n\mathbf{k}}}{4\xi_{n\mathbf{k}}^2}
+ \frac{\mathrm{i} \pi}{8 T_c N} \sum_n D_n(\mu) \theta_\mu,
\label{eqn:d}
\end{align}
where
$
\mathcal{Y}_{n\mathbf{k}} = \mathrm{sech^2}\big(\frac{\xi_{n\mathbf{k}}}{2T}\big)
$
is a thermal factor and 
$
\dot{\xi}_{n\mathbf{k}}^i = \partial \xi_{n\mathbf{k}} / \partial k_i.
$
In order to reproduce the correct $U_{pp}$ that is previously obtained 
from the low-energy collective-mode analysis at $T = 0$~\cite{iskin24}, 
i.e., see Sec.~\ref{sec:num} for further discussion, here the expansion 
coefficients are given per lattice site. We note that all of these 
coefficients must be evaluated at $T_c$ self-consistently with $\mu$. 
In addition, since the factor
$
\frac{\partial \mu}{\partial T}
$  
plays a crucial role away from the BCS limit, Eq.~(\ref{eqn:a0}) has 
to be handled with care in flat-band superconductors, which is 
described in Sec.~\ref{sec:num}. However, numerical implementation of 
the rest of the coefficients is a straightforward task once $(\mu, T_c)$ 
is computed for a desired $U$. 

Motivated by the recent surge of interest in the quantum-geometric effects, 
here we also split the kinetic coefficient into two contributions
$
c_{ij} = c_{ij}^\mathrm{intra} + c_{ij}^\mathrm{inter},
$
depending on whether the intraband or interband processes are involved. 
Note that, in comparison to our previous work~\cite{iskin23}, here we 
used integration by parts 
$
\sum_\mathbf{k} \mathcal{X}_{n\mathbf{k}} \ddot{\xi}_{n\mathbf{k}}^{ij} / \xi_{n\mathbf{k}}^2
= \sum_\mathbf{k} \big[ 2 \mathcal{X}_{n\mathbf{k}} / \xi_{n\mathbf{k}}^3
- \mathcal{Y}_{n\mathbf{k}} / (2 T \xi_{n\mathbf{k}}^2) \big]
\dot{\xi}_{n\mathbf{k}}^i \dot{\xi}_{n\mathbf{k}}^j
$
and
$\sum_\mathbf{k} \mathcal{Y}_{n\mathbf{k}} \ddot{\xi}_{n\mathbf{k}}^{ij} / \xi_{n\mathbf{k}}
=
\sum_\mathbf{k} \big[ \mathcal{Y}_{n\mathbf{k}} / \xi_{n\mathbf{k}}^2
+ \mathcal{X}_{n\mathbf{k}} \mathcal{Y}_{n\mathbf{k}} / (T \xi_{n\mathbf{k}}) \big]
\dot{\xi}_{n\mathbf{k}}^i \dot{\xi}_{n\mathbf{k}}^j,
$
and reexpressed $c_{ij}^\mathrm{intra}$ in the new form given by 
Eq.~(\ref{eqn:cintra}), where
$
\ddot{\xi}_{n\mathbf{k}}^{ij} = \partial^2 \xi_{n\mathbf{k}} 
/ \partial k_i \partial k_j.
$
This coefficient is the so-called conventional contribution, as it is 
simply a sum over its single-band counterpart~\cite{iskin23}.
Similarly, the coefficients $a_0$, $b_0$ and $d$ are sums over their 
single-band counterparts.
In contrast, the geometric contribution $c_{ij}^\mathrm{inter}$ is 
controlled by the quantum-metric tensor 
$
g_{ij}^{n\mathbf{k}}  = \sum_{n\ne m} g_{ij}^{n m\mathbf{k}}
$
of the $n$th Bloch band~\cite{Provost80, berry84, resta11} and 
its band-resolved quantum-metric tensor 
$
g_{ij}^{n m\mathbf{k}} = 2\mathrm{Re} \left[ 
\langle \dot{n}_\mathbf{k}^i | m_\mathbf{k} \rangle
\langle m_\mathbf{k} | \dot{n}_\mathbf{k}^j \rangle
\right],
$
where $\mathrm{Re}$ denotes the real part and 
$
| \dot{n}_\mathbf{k}^i \rangle = \partial |n_{\mathbf{k}} \rangle/\partial k_i.
$
Such terms follow from the low-$\mathbf{q}$ expansion of the Bloch factor 
\begin{align}
|\langle n_\mathbf{k} | m_\mathbf{k-q} \rangle|^2 
= \delta_{nm} - \frac{1}{2} \sum_{ij} [g_{ij}^{n\mathbf{k}} \delta_{nm} 
+ g_{ij}^{nm\mathbf{k}}(\delta_{nm} - 1)] q_i q_j
\label{eqn:gnm}
\end{align}
that appears in Eq.~(\ref{eqn:Gamma0}). It is important to highlight that 
$c_{ij}^\mathrm{inter}$ does not have any contribution from the band touchings, 
i.e., the first term of Eq.~(\ref{eqn:cinter}) cancels those touching 
contributions from the second term whenever $\xi_{n\mathbf{k}} = \xi_{m\mathbf{k}}$ 
for any $n \ne m$. For instance, in the case of pyrochlore lattice, there is 
no inter-flat-band contribution to $c_{ij}^\mathrm{inter}$ among the degenerate 
flat bands as they touch each other everywhere in the BZ.

Furthermore, we note that the dynamic coefficient $d$ is a complex number 
in general~\cite{iskin23, sademelo93}, and its imaginary part is determined 
by the density of states 
$
D_n(\varepsilon) = \sum_\mathbf{k} \delta(\varepsilon - \varepsilon_{n\mathbf{k}})
$
of the $n$th Bloch band where
$
\theta_\varepsilon = \theta(\varepsilon - \min\{\varepsilon_{n\mathbf{k}}\})
\theta(\max\{\varepsilon_{n\mathbf{k}}\} - \varepsilon).
$
Here, $\delta(x)$ is the Dirac-delta distribution and $\theta(x)$ is the 
Heaviside-step function. Thus, $d$ has a dominant positive imaginary part 
in the BCS limit when $\mu$ lies within any one of the Bloch bands. 
This is nothing but a reflection of the continuum of fermionic excitations 
into which a Cooper pair can decay, suggesting that the dynamics of 
the order parameter is overdamped in the BCS limit. In sharp contrast, 
$d$ becomes purely real in the BEC limit suggesting that the dynamics of 
the order parameter is propagating. For the dilute flat-band superconductor 
of interest in this paper, the latter turns out to the case for any $U \ne 0$.

Assuming this is the case, and by making an analogy with the Gross-Pitaevskii 
equation for a weakly-interacting atomic Bose gas, i.e., through a scaling of 
the order parameter~\cite{iskin23, sademelo93}, we define
$
\mu_p(T) = a_0 \epsilon (T) / d
$
as the effective chemical potential of the Cooper pairs,
$
(M_p^{-1})_{ij} = c_{ij} / d
$
as the inverse effective-mass tensor $\mathbf{M}_p^{-1}$ of the pairs, and
$
U_{pp} = b_0 / d^2
$
as the effective onsite repulsive interaction between the pairs. 
This identification suggests that $\mathbf{M}_p$ is composed of a 
conventional contribution 
$
(M_p^{-1})^\mathrm{intra}_{ij} = c_{ij}^\mathrm{intra} / d
$
and a geometric contribution
$
(M_p^{-1})^\mathrm{inter}_{ij} = c_{ij}^\mathrm{inter} / d.
$
In this paper, we are primarily interested in the GL coherence length
$
(\xi_T^2)_{ij} = (\xi_\mathrm{GL}^2)_{ij} / \epsilon (T),
$
whose temperature-independent prefactor is given by
\begin{align}
(\xi_\mathrm{GL}^2)_{ij} &= \frac{c_{ij}}{2 a_0}.
\label{eqn:GL}
\end{align}
This relation again suggests that $\xi_\mathrm{GL}$ is composed of a 
conventional contribution 
$
(\xi_\mathrm{GL}^2)^\mathrm{intra}_{ij} = c_{ij}^\mathrm{intra}/(2 a_0)
$
and a geometric contribution
$
(\xi_\mathrm{GL}^2)^\mathrm{inter}_{ij} = c_{ij}^\mathrm{inter}/(2 a_0).
$
Since the calculation of $\xi_\mathrm{GL}$ requires $\mu$ and $T_c$ as 
inputs, we next describe our recipe for their self-consistent evaluation 
as a function of $U$, which is appropriate for a dilute flat-band 
superconductor in the pyrochlore lattice.

\section{Self-consistency relations: $(\mu, T_c)$}
\label{sec:sce}

As the simplest route, we follow the usual finite-temperature BCS-BEC 
crossover formalism~\cite{sademelo93, nsr85} to construct a self-consistent 
theory based on the Thouless condition, which is equivalent to both 
the saddle-point condition and the mean-field order-parameter equation, 
and the number equation
$
\mathcal{N} = - \partial \Omega / \partial \mu.
$
Here, $\Omega$ is the thermodynamic potential, and keeping its 
corrections at the Gaussian order, i.e., 
$
\Omega_\mathrm{G} = \Omega_0 + \Omega_2,
$ 
is known to be sufficient in producing a qualitatively correct physical 
description of the system for all $U \ne 0$. 

One can show quite generally that
$
\Omega_0 = T \mathcal{S}_0
$
is the saddle-point and
$
\Omega_2 = T \sum_q \ln \det [T \boldsymbol{\Gamma}^{-1}(q)]
$
is the quadratic contribution. At $T_c$, while the former leads to the 
Fermi-Dirac (FD) distribution of an unbound (free) Fermi gas of $\uparrow$ 
and $\downarrow$ fermions, the latter contribution 
$
\mathcal{N}_2 = \sum_\mathbf{q} \int_{-\infty}^{+\infty} \frac{d\omega}{\pi} 
f_\mathrm{BE}(\omega) \frac{\partial \delta(\mathbf{q},\omega)}{\partial \mu},
$
where
$
\delta(\mathbf{q},\omega) = -\mathrm{Arg}[\det \boldsymbol{\Gamma}^{-1}(\mathbf{q}, \omega+\mathrm{i}0^+)]
$
is the argument of the propagator, is typically split into
$
\mathcal{N}_2 = \mathcal{N}_{bs} + \mathcal{N}_{sc}
$
~\cite{sademelo93, nsr85}.
While the bound-state contribution $\mathcal{N}_{bs}$ arises from the 
isolated poles of $\boldsymbol{\Gamma}(q)$, the continuum of two-particle 
excitations leads to the scattering contribution $\mathcal{N}_{sc}$ 
arising from the branch cut of the logarithm. Suppose the branch point
is at $\omega = \omega_\mathbf{q}^*$ below which $\delta(\mathbf{q},\omega) = 0$.
Thus, 
$
\mathcal{N}_{bs} = -T \sum_q \frac{\partial 
[\det \boldsymbol{\Gamma}^{-1}(q)]/\partial \mu}
{\det \boldsymbol{\Gamma}^{-1}(q)}
$
can be simplified considerably by noting that
$
\det \boldsymbol{\Gamma}^{-1}(q) \propto 
\Pi_s (\mathrm{i}\nu_\ell - \omega_{s\mathbf{q}}),
$
where $0 \le \omega_{s\mathbf{q}} < \omega_\mathbf{q}^* - 2\mu$ are the 
poles of $\boldsymbol{\Gamma}(q)$ determined by 
$
\det \boldsymbol{\Gamma}^{-1}(\mathbf{q}, \omega_{s\mathbf{q}}) = 0.
$
For instance, since 
$
\partial \omega_{s\mathbf{q}}/\partial \mu = -2
$
in the BEC limit where the bound state is made of two fermions with 
opposite spins, one can approximate
$
\mathcal{N}_{bs} = -T \sum_{\ell s \mathbf{q}} 2/(\mathrm{i}\nu_\ell - \omega_{s\mathbf{q}})
= 2\sum_{s \mathbf{q}} f_\mathrm{BE}(\omega_{s \mathbf{q}}),
$
after summing over the Matsubara frequencies, where
$
f_\mathrm{BE}(x) = 1/(e^{x/T} - 1) = [\coth\big(\frac{x}{2T}\big) - 1]/2
$
is the Bose-Einstein (BE) distribution of a bound Bose gas of Cooper pairs.
Here, the $\mathbf{q}$ sum is restricted to only those center-of-mass 
momenta for which the corresponding bound states exist.
This analysis is equivalent to approximating
$
\delta_{bs}(\mathbf{q},\omega) = \pi \sum_s \theta(\omega - \omega_{s\mathbf{q}})
$
as the bound-state contribution to the phase shift near 
$
\omega \approx \omega_{s\mathbf{q}},
$ 
where $\theta(x)$ is the Heaviside step function~\cite{nsr85, sademelo93}.
On the other hand, the branch-cut contribution is given by
$
\mathcal{N}_{sc} = \sum_\mathbf{q} \int_{\omega_\mathbf{q}^*}^{+\infty} 
\frac{d\omega}{\pi} 
f_\mathrm{BE}(\omega) \frac{\partial \delta(\mathbf{q},\omega)}{\partial \mu}.
$
For instance, if we decompose
$
\det \boldsymbol{\Gamma}^{-1}(\mathbf{q}, \omega+\mathrm{i}0^+) 
= R + \mathrm{i} I
$
into its real and imaginary parts then
$
\frac{\partial \delta(\mathbf{q},\omega)}{\partial \mu} 
= (\dot{R}^\mu I - R \dot{I}^\mu)/(R^2 + I^2),
$
where 
$
\dot{R}^\mu = \partial R/\partial \mu
$
and
$
\dot{I}^\mu = \partial I/\partial \mu.
$
Since accurately computing $\mathcal{N}_{sc}$ is quite challenging, 
even in the context of much simpler continuum 
problems~\cite{sademelo93, nsr85}, we next focus only on a dilute 
flat-band superconductor where $\mathcal{N}_{sc}$ is negligible 
for all $U \ne 0$, and can be omitted.

Thus, having a dilute flat-band superconductor in the pyrochlore lattice 
in mind, our self-consistency relations for $\mu$ and $T_c$ can be 
summarized as
\begin{align}
\label{eqn:Tc}
1 &= \frac{U}{N} \sum_{n\mathbf{k}} \frac{\mathcal{X}_{n\mathbf{k}}}{2\xi_{n\mathbf{k}}}, \\
F &= \frac{2}{N} \sum_{n\mathbf{k}} f_\mathrm{FD}(\xi_{n\mathbf{k}}) 
+ \frac{2}{N} \sum_{\mathbf{q}} f_\mathrm{BE}(\omega_{b\mathbf{q}}),
\label{eqn:F}
\end{align}
where 
$
0 \le F = \mathcal{N}/N \le 2
$
is the average filling of particles per lattice site and
$
f_\mathrm{FD}(x) = 1/(e^{x/T} + 1) = [1 - \tanh\big(\frac{x}{2T}\big)]/2
$
is the FD distribution. In Eq.~(\ref{eqn:F}), the prefactor of 2 in front 
of the FD distribution is due to spin degeneracy, while that of the 
BE distribution can be traced back to the presence of 2 particles 
in a two-body bound state. Since the flat bands appear at the 
bottom of the Bloch spectrum in the pyrochlore lattice, we emphasize that
Eq.~(\ref{eqn:F}) is very accurate in the $F \ll 1$ limit for all $U \ne 0$, 
but it may still give qualitatively correct results up to the half-filling 
for the flat bands, i.e., up to $F \lesssim 0.5$. In addition, it is, 
by construction, accurate for other multiband lattices up to their half fillings, 
i.e., up to $F \lesssim 1$, in the $U/t \gg 1$ limit. In Eq.~(\ref{eqn:F}), 
it is sufficient to keep only one of the pole contributions, i.e., 
the one with the lowest energy, because poles with higher energies are 
expected to give negligible contributions due to the BE distribution. 
Note that, in the presence of 
time-reversal symmetry and uniform pairing, this is also equivalent to 
keeping the isolated pole of Eq.~(\ref{eqn:Gamma0}). Furthermore, in our 
numerical calculations, we implement the simplest approach and extract 
$\omega_{b\mathbf{q}}$ from the dispersion $E_{1\mathbf{q}}$ of the 
lowest-lying two-body bound-state branch as follows.

In general, all of the two-body bound states can be determined from the 
isolated poles of Eq.~(\ref{eqn:GammaSS}) exactly, i.e., by computing
$
\det \boldsymbol{\Gamma}^{-1}(\mathbf{q}, E_{s\mathbf{q}}) = 0,
$
after setting the thermal factors
$
\mathcal{X}_{n\mathbf{k}\sigma} \to 1
$
and $\mu \to 0$~\cite{iskin23}. However, in the presence of time-reversal 
symmetry and uniform pairing, the lowest-lying branch can simply be 
determined from the isolated pole of Eq.~(\ref{eqn:Gamma0}), 
i.e., by computing
$
\Gamma_0^{-1}(\mathbf{q}, E_{1\mathbf{q}}) = 0,
$
after again setting the thermal factors
$
\mathcal{X}_{n\mathbf{k}} \to 1
$
and $\mu \to 0$. It can be shown that the overall structure 
of the two-body bound-state branches $E_{s\mathbf{q}}$ resemble the 
underlying Bloch bands $\varepsilon_{n\mathbf{k}}$ but with the opposite 
sign of energy~\cite{iskin24}. 
For instance, $E_{1\mathbf{q}}$ resembles to $-\varepsilon_{1\mathbf{k}}$ 
but with an effective pair hopping parameter $t_b$ and some energy offset 
depending on $U$. To determine these effective parameters, we first expand
$
\varepsilon_{1\mathbf{k}} = -6\bar{t} + a^2 \bar{t} |\mathbf{k}|^2/8
$
in the low $\mathbf{k} \to \mathbf{0}$ limit, and identify the relation 
between the hopping parameter and the effective mass $m_1$ of the unpaired 
spin-$\sigma$ particles as $\bar{t} = 4/(a^2 m_1)$
~\footnote{
We note that the matrix elements
$
(m_n^{-1})_{ij} = \ddot{\varepsilon}_{n \mathbf{k}_0}^{ij}
$ 
of the inverse effective band-mass tensor $\mathbf{m}_n^{-1}$ of the $n$th 
Bloch band itself can also be written as
$
(m_n^{-1})_{ij} = (m_n^{-1})_{ij}^\mathrm{intra} + (m_n^{-1})_{ij}^\mathrm{inter},
$
where 
$
(m_n^{-1})_{ij}^\mathrm{intra} = \langle n_{\mathbf{k}_0}| 
\ddot{\mathbf{h}}_{\mathbf{k}_0}^{ij} |n_{\mathbf{k}_0} \rangle
$
is the conventional intraband contribution and 
$
(m_n^{-1})_{ij}^\mathrm{inter} = \sum_m 
(\varepsilon_{n \mathbf{k}_0} - \varepsilon_{m \mathbf{k}_0})
g_{ij}^{n m \mathbf{k}_0}
$
is the geometric interband contribution~\cite{iskin19b}. Here, $\mathbf{k}_0$ 
denotes the location of the band minimum or maximum of the corresponding 
Bloch band, and 
$
\ddot{\mathbf{h}}_{\mathbf{k}_0}^{ij} = \frac{\partial^2 \mathbf{h}_\mathbf{k}}
{\partial k_i \partial k_j}\big|_{\mathbf{k}_0}
$
}.
This suggests that 
$
t_b = 4/(a^2 M_b)
$
is the relation between the effective hopping parameter and effective mass 
$M_b$ of the lowest-lying two-body bound states. 

In the low-$\mathbf{q}$ limit, we also recall that
$
E_{1 \mathbf{q}} = E_b + \frac{1}{2}\sum_{ij} (M_b^{-1})_{ij} q_i q_j
$
~\cite{iskin22}, where the energy offset $E_b$ is determined by
$
1 = \frac{U}{N} \sum_{n \mathbf{k}} 1/(2\varepsilon_{n \mathbf{k}} - E_b).
$
Furthermore, very similar to that of the many-body bound states 
discussed in Sec.~\ref{sec:coeff}, the inverse effective-mass tensor
$
(M_b^{-1})_{ij} = (M_b^{-1})^\mathrm{intra}_{ij} + (M_b^{-1})^\mathrm{inter}_{ij}
$
of the lowest-lying two-body bound states is composed of an analogous 
conventional contribution
$
(M_b^{-1})^\mathrm{intra}_{ij} = \frac{1}{2D}\sum_{n \mathbf{k}} \ddot{\varepsilon}_{n\mathbf{k}}^{ij}
/(2\varepsilon_{n \mathbf{k}} - E_b)^2
$
and an analogous geometric one
$
(M_b^{-1})^\mathrm{inter}_{ij} = \frac{1}{D}
\sum_{n \mathbf{k}} g_{ij}^{n \mathbf{k}}/(2\varepsilon_{n \mathbf{k}} - E_b)
- \frac{1}{D}\sum_{n, m \ne n, \mathbf{k}} g_{ij}^{nm\mathbf{k}}/
(\varepsilon_{n \mathbf{k}} + \varepsilon_{m \mathbf{k}} - E_b).
$
Using integration by parts, i.e.,
$
\sum_{\mathbf{k}} \ddot{\varepsilon}_{n\mathbf{k}}^{ij}
/(2\varepsilon_{n \mathbf{k}} - E_b)^2 = 
4\sum_{\mathbf{k}} \dot{\varepsilon}_{n\mathbf{k}}^{i} \dot{\varepsilon}_{n\mathbf{k}}^{j}
/(2\varepsilon_{n \mathbf{k}} - E_b)^3,
$
the former expression can alternatively be written in the form of 
Eq.~(\ref{eqn:cintra}). In addition, similar to $c_{ij}^\mathrm{inter}$, it is 
pleasing to note that $(M_b^{-1})^\mathrm{inter}_{ij}$ also does not have any contribution 
from the band touchings, i.e., the first sum cancels those touching contributions 
from the second sum whenever $\xi_{n\mathbf{k}} = \xi_{m\mathbf{k}}$ for any 
$n \ne m$. It is also worth emphasizing that these two-body expressions are 
exact for any $U \ne 0$ in the pyrochlore lattice~\cite{iskin24}, and
$
(M_b^{-1})_{ij} = \delta_{ij}/M_b
$
is diagonal as a consequence of uniform pairing. 
Thus, $E_b$ and $M_b$ can be determined self-consistently through these
analytical expressions as functions of $U$, where
$
D = \sum_{n \mathbf{k}} 1/(2\varepsilon_{n \mathbf{k}} - E_b)^2.
$
Once $E_b$ and $M_b$ are determined, we identify that
\begin{align}
\omega_{b\mathbf{q}} = E_{1\mathbf{q}} - 2\mu = \varepsilon_{b \mathbf{q}} - \mu_b,
\label{eqn:wbq}
\end{align}
leading to
$
\varepsilon_{b \mathbf{q}} = 6t_b - 2t_b(1 + \sqrt{1 + \alpha_\mathbf{q}})
$
as the effective dispersion, where $\alpha_\mathbf{q}$ is defined in
Sec.~\ref{sec:Ham}, and
$
\mu_b = 2\mu - E_b \to 0^-
$
as the effective chemical potential of pairs, such that
$
\varepsilon_{b \mathbf{q}} = |\mathbf{q}|^2/(2M_b)
$
reproduces the two-body physics exactly for all $U \ne 0$ in the low 
$\mathbf{q} \to \mathbf{0}$ limit. In Sec.~\ref{sec:num}, we benchmark this 
approximate yet relatively complicated finite-temperature recipe against a 
much simpler zero-temperature one for the pyrochlore lattice, demonstrating 
very good agreement between the two.

\section{Coherence length at $T = 0$}
\label{sec:T0}

To compare with the GL coherence length $\xi_\mathrm{GL}$, here we derive the
so-called zero-temperature coherence length $\xi_0$~\cite{pistolesi96}, 
based on the effective Gaussian action for the order-parameter 
fluctuations~\cite{iskin24}. Using the Grassmann functional-integral 
formalism, and under the same assumptions discussed above, one can show that
$
\mathcal{S}_2 = \frac{N}{2T} \sum_q \left( \Lambda_q^* \, \Lambda_{-q} \right)
\boldsymbol{\mathcal{M}}^q
\left( \begin{array}{c}
\Lambda_q \\ \Lambda_{-q}^*
\end{array} \right)
$
is the quadratic action, where the matrix $\boldsymbol{\mathcal{M}}^q$ plays 
the role of an inverse propagator for the fluctuations. The explicit 
finite-temperature expressions for its matrix elements, 
$
\mathcal{M}^q_{11} = \mathcal{M}^{-q}_{22}
$
and
$
\mathcal{M}^q_{12} = \mathcal{M}^q_{21},
$
can be found in our previous work~\cite{iskin24}. 
The off-diagonal terms vanish as $T \to T_c$, and the diagonal ones give 
rise to $\Gamma_0^{-1}(q)$.

We define $\xi_0$ by setting $\mathrm{i} \nu_\ell = 0$ and expanding 
$
\mathcal{M}^\mathbf{q}_{11} + \mathcal{M}^\mathbf{q}_{12}
$
at $T = 0$ up to second order in $\mathbf{q}$, corresponding to the 
amplitude-amplitude matrix element of the fluctuation propagator in the 
long-wavelength limit~\cite{pistolesi96}. This leads to
$
\mathcal{M}^\mathbf{q}_{11} + \mathcal{M}^\mathbf{q}_{12} 
= A + \sum_{ij} C_{ij} q_i q_j
$
~\cite{iskin24}, where
\begin{align}
\label{eqn:A}
A &= \frac{1}{N} \sum_{n \mathbf{k}} \frac{\Delta_0^2}{2E_{n \mathbf{k}}^3}, \\
\label{eqn:Cintra}
C_{ij}^\mathrm{intra} &= \frac{1}{N} \sum_{n \mathbf{k}} 
\frac{1}{8 E_{n\mathbf{k}}^3} 
\left( 1 - \frac{5\Delta_0^2 \xi_{n\mathbf{k}}^2}{E_{n\mathbf{k}}^4} \right)
\dot{\xi}_{n\mathbf{k}}^i \dot{\xi}_{n\mathbf{k}}^j, \\
C_{ij}^\mathrm{inter} &= 
\frac{1}{N} \sum_{n \mathbf{k}} 
\frac{\xi_{n\mathbf{k}}^2} {4 E_{n\mathbf{k}}^3} g^{n\mathbf{k}}_{ij} \nonumber \\
&- \frac{1}{N} \sum_{n, m\ne n, \mathbf{k}} 
\frac{\xi_{n\mathbf{k}}\xi_{m\mathbf{k}} + E_{n\mathbf{k}} E_{m\mathbf{k}} - \Delta_0^2}
{4 E_{n\mathbf{k}} E_{m\mathbf{k}} (E_{n\mathbf{k}}+E_{m\mathbf{k}})}
g^{nm\mathbf{k}}_{ij},
\label{eqn:Cinter}
\end{align}
Here, $\Delta_0$ is the saddle-point, i.e., the mean-field, order parameter 
for pairing, and 
$
E_{n\mathbf{k}} = \sqrt{\Delta_0^2 + \xi_{n\mathbf{k}}^2}
$
is the quasi-particle energy associated with the $n$th Bloch band. In order 
to make a direct comparison with $a_0$ and $c_{ij}$ discussed in 
Sec.~\ref{sec:coeff}, these expansion coefficients are also given per lattice 
site. Furthermore, we again split the kinetic coefficient into two contributions
$
C_{ij} = C_{ij}^\mathrm{intra} + C_{ij}^\mathrm{inter},
$
depending on whether the intraband or interband processes are involved. 
This is in such a that the coefficients $A$ and $C_{ij}^\mathrm{intra}$ 
are again simply sums over their conventional single-band 
counterparts~\cite{iskin24, engelbrecht97, pistolesi96}. 
The latter can be verified through the integration by parts
$
\sum_\mathbf{k} \xi_{n \mathbf{k}} \ddot{\xi}_{n \mathbf{k}}^{ij}/E_{n \mathbf{k}}^3
= \sum_\mathbf{k} (3\xi_{n \mathbf{k}}^2 - E_{n \mathbf{k}}^2) 
\dot{\xi}_{n \mathbf{k}}^i \dot{\xi}_{n \mathbf{k}}^j/E_{n \mathbf{k}}^5
$
and
$
\sum_\mathbf{k} \xi_{n \mathbf{k}} \ddot{\xi}_{n \mathbf{k}}^{ij}/E_{n \mathbf{k}}^5
= \sum_\mathbf{k} (5\xi_{n \mathbf{k}}^2 - E_{n \mathbf{k}}^2) 
\dot{\xi}_{n \mathbf{k}}^i \dot{\xi}_{n \mathbf{k}}^j/E_{n \mathbf{k}}^7
$
in Eq.~(\ref{eqn:Cintra}). Aside from a typo in the first line of Eq.~(28) 
in our previous work~\cite{iskin24}, i.e., $E_{n \mathbf{k}}$ is supposed 
to be $E_{n \mathbf{k}}^3$ in the denominator,
Eq.~(\ref{eqn:Cinter}) proves to be a more convenient alternative expression 
for $C_{ij}^\mathrm{inter}$ when considering the dilute limit below.
Similar to $c_{ij}^\mathrm{inter}$ and $(M_b^{-1})^\mathrm{inter}_{ij}$, it is 
pleasing to note that $C_{ij}^\mathrm{inter}$ also does not have any contribution 
from the band touchings, i.e., the first term of Eq.~(\ref{eqn:Cinter}) cancels 
those touching contributions from the second term whenever 
$\xi_{n\mathbf{k}} = \xi_{m\mathbf{k}}$ for any $n \ne m$. 

In accordance with the literature~\cite{pistolesi96}, we define the 
zero-temperature coherence length as
\begin{align}
(\xi_\mathrm{0}^2)_{ij} &= \frac{C_{ij}}{A},
\label{eqn:xi0}
\end{align}
assuming $C_{ij} > 0$. Here, $C_{ij} < 0$ signals that the minimum of 
$
\mathcal{M}^\mathbf{q}_{11} + \mathcal{M}^\mathbf{q}_{12}
$
occurs at a finite $\mathbf{q}$~\cite{benfatto02}, depending on the 
symmetries of the underlying lattice geometry. In such cases, the 
low-$\mathbf{q}$ expansion must be performed around the new minimum 
instead of $\mathbf{q = 0}$. Similar to $\xi_\mathrm{GL}$, 
Eq.~(\ref{eqn:xi0}) suggests that $\xi_0$ is also composed of a 
conventional contribution 
$
(\xi_0^2)^\mathrm{intra}_{ij} = C_{ij}^\mathrm{intra}/A
$
and a geometric one
$
(\xi_0^2)^\mathrm{inter}_{ij} = C_{ij}^\mathrm{inter}/A.
$
Here, the calculation of $\xi_0$ requires self-consistent solutions of $\mu$ 
and $\Delta_0$ as inputs. In the usual zero-temperature BCS-BEC crossover 
formalism~\cite{engelbrecht97, iskin24}, these parameters are determined by 
the saddle-point condition
$
1 = \frac{U}{N} \sum_{n\mathbf{k}} 1/(2E_{n\mathbf{k}})
$
and the mean-field number equation
$
F = F_0 = 1 - \frac{1}{N} \sum_{n\mathbf{k}} \xi_{n\mathbf{k}}/ E_{n\mathbf{k}}.
$
In sharp contrast to the finite-temperature recipe of Sec.~\ref{sec:sce}, 
which requires Gaussian corrections to the number equation, this simple 
mean-field recipe is known to be sufficient in producing a qualitatively 
correct physical description of the system for all $U \ne 0$. In fact,
in the case of a three-dimensional continuum model with a quadratic 
dispersion, $\xi_0$ and $\xi_\mathrm{GL}$ are known to give very similar 
results up to a constant of order unity for all $U \ne 0$~\cite{pistolesi96}. 
However, this is not in general the case for a single-band lattice model 
with cosine dispersion away from the dilute limit, where $C_{ij}$ can 
change sign and become negative for some intermediate $U$ values~\cite{benfatto02}.
See Appendix~\ref{sec:appA} for such a breakdown in the pyrochlore lattice.

Before delving into the heavy numerical calculations, here we 
highlight an important analytical insight for the pyrochlore lattice. 
In the dilute flat-band limit when $F \ll 1$, one can show that 
$\mu + 2t < 0$ and $|\mu+2t| \gg \Delta_0$ for all $U \ne 0$. 
For instance, this can be verified by taking the $F \to 0$ limit of
$
\Delta_0 = \frac{U}{2} \sqrt{F(1-F)}
$ 
and
$
\mu = - 2t - \frac{U}{4}(1-2F)
$
in the $U/t \ll 1$ limit, and of
$
\Delta_0 = \frac{U}{2} \sqrt{F(2-F)}
$ 
and
$
\mu = - \frac{U}{2}(1-F)
$
in the $U/t \gg 1$ limit. Thus, since 
$
\xi_{n \mathbf{k}} \gg \Delta_0
$ 
for any $U \ne 0$, we may set 
$
E_{n \mathbf{k}} \to \xi_{n \mathbf{k}},
$
leading to
$
A = \frac{1}{N} \sum_{n \mathbf{k}} \Delta_0^2/(2 \xi_{n\mathbf{k}}^3)
$
for the static coefficient,
$
C_{ij}^\mathrm{intra} = \frac{1}{N} \sum_{n \mathbf{k}} 
\dot{\xi}_{n\mathbf{k}}^i \dot{\xi}_{n\mathbf{k}}^j/(8 \xi_{n\mathbf{k}}^3)
$
for the conventional kinetic coefficient, and
$
C_{ij}^\mathrm{inter} = \frac{1}{2N} \sum_{n \mathbf{k}}
g^{n\mathbf{k}}_{ij}/(2\xi_{n\mathbf{k}})
-  \frac{1}{2N} \sum_{n, m \ne n, \mathbf{k}} 
g^{nm\mathbf{k}}_{ij}/(\xi_{n\mathbf{k}}+\xi_{m\mathbf{k}})
$
for the geometric kinetic coefficient. 
Similarly, since $\xi_{n \mathbf{k}} \gg T_c$ for any 
$U \ne 0$ in the $F \ll 1$ limit, we may set $\mathcal{X}_{n \mathbf{k}} \to 1$ 
and $\mathcal{Y}_{n \mathbf{k}} \to 0$ in Eqs.~(\ref{eqn:a0}),~(\ref{eqn:cintra})
and~(\ref{eqn:cinter}), leading to 
$
a_0 = -\frac{T_c}{2N} \frac{\partial \mu}{\partial T}\big{|}_{T_c} 
\sum_{n \mathbf{k}} 1/\xi_{n \mathbf{k}}^2
$
for the static coefficient, and 
$
c_{ij}^\mathrm{intra} = 2C_{ij}^\mathrm{intra}
$ 
and
$
c_{ij}^\mathrm{inter} = 2C_{ij}^\mathrm{inter}
$ 
for the kinetic coefficients.
Note that a prefactor of 2 difference arises from the distinct definitions 
used in the low-$\mathbf{q}$ expansions of $\Gamma_0^{-1}(\mathbf{q})$ and 
$
\mathcal{M}^\mathbf{q}_{11} + \mathcal{M}^\mathbf{q}_{12}.
$
Thus, in the dilute limit, we find that 
$
(\xi_0^2)_{ij}/(\xi_\mathrm{GL}^2)_{ij} = a_0/A
$
for all $U \ne 0$, assuming these length scales are nonzero.

\section{Numerical Results}
\label{sec:num}

\begin{figure} [htb]
\includegraphics[width = 0.99\linewidth]{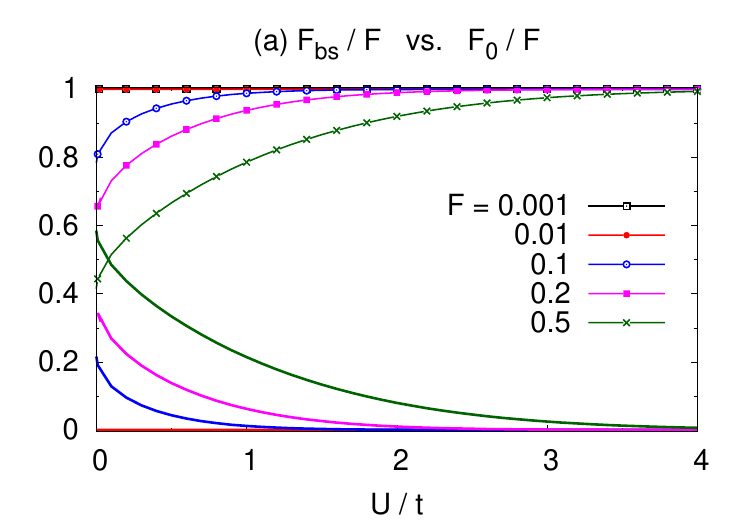}
\includegraphics[width = 0.99\linewidth]{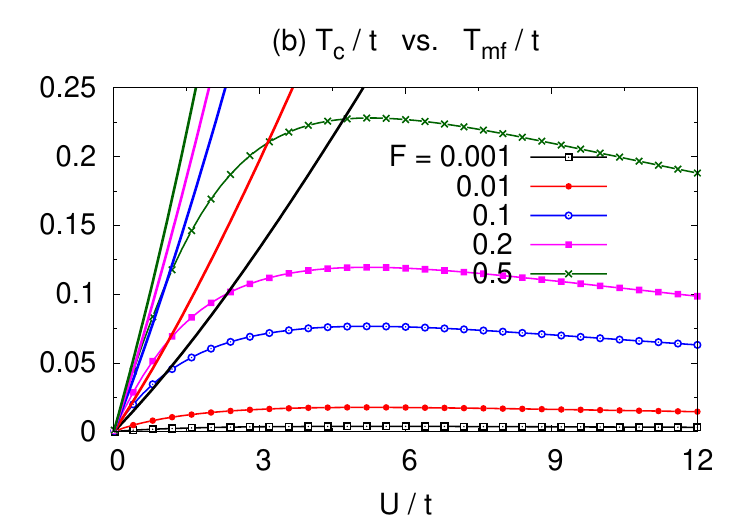}
\caption{\label{fig:Tc}
Fractions of bound-state contribution (line plots with data points) and 
saddle-point contribution (line plots) are shown in (a) as functions 
of interaction strength for various fillings. The data for fillings 
$F = 0.01$ and $F = 0.001$ are indistinguishable on the presented scale. 
Corresponding critical temperatures (line plots with data points) are 
shown in (b), and compared with the mean-field temperature scales 
(line plots).
}
\end{figure}

In Fig.~\ref{fig:Tc}, we present self-consistent solutions of Eqs.~(\ref{eqn:Tc})
and~(\ref{eqn:F}) as functions of $U/t$ for various particle fillings. 
Here, we recall that the scattering-state contribution $F_{sc}$ to the number 
equation is not included in our simplified recipe, i.e., $F = F_0 + F_{bs}$. 
Therefore, our numerical results are expected to be accurate in parameter 
regimes where either the saddle-point or the bound-state contribution 
dominate. This occurs when either
$
F_{0}/F \to 1
$
leading to the BCS limit, or
$
F_{bs}/F \to 1
$
leading to the BEC limit.
Figure~\ref{fig:Tc}(a) suggests that our approximations are well-justified 
for all $U/t \ne 0$ for a dilute flat-band superconductor with $F \ll 1$, 
and for $U/t \gg 1$ in general for higher fillings. In this paper, we are 
not concerned with those fillings where $\mu$ overlaps with a dispersive 
band in the non-interacting limit. In such cases, similar to the usual BCS-BEC 
crossover problem~\cite{sademelo93, nsr85}, our recipe can also be justified 
in both the low-$U/t$ and high-$U/t$ regimes, but not in the intermediate 
crossover region.

The corresponding $T_c$ data are presented in Fig.~\ref{fig:Tc}(b). 
We observe that after $T_c$ rises almost linearly 
for small $U \lesssim t$, it reaches a peak around $U \sim 5t$ and then 
decays as $t^2/U$ for $U \gg t$. These results are consistent with recent 
works on the Berezinskii-Kosterlitz-Thouless (BKT) transition temperature 
$T_\mathrm{BKT}$ of two-dimensional flat-band models, which are based on the 
universal-jump relation involving the superfluid-weight 
tensor~\cite{torma22, huhtinen22}.
In comparison, we also present the mean-field temperature scale $T_{mf}$,
which is calculated by setting $F_{bs} = 0$ in Eq.~(\ref{eqn:F}). 
Unlike $T_c$, which marks the onset of phase coherence among Cooper pairs, 
$T_{mf}$ is associated with the formation of pairs and increases 
without bound as a function of $U/t$~\cite{sademelo93, nsr85}. 
In other words, the dissociation temperature of pairs increases with 
their binding energy. Thus, in a flat-band superconductor, 
Fig.~\ref{fig:Tc}(b) shows that the formation and condensation of Cooper 
pairs occur at very different temperature scales for any $U \ne 0$, 
and the BCS theory must be used with caution. While $T_c$ and $T_\mathrm{BKT}$
are governed directly by the quantum geometry of the Bloch states through 
$F_{bs}$ and superfluid weight, respectively, in three and two dimensions, 
$T_{mf}$ is not.

After computing the self-consistent solutions for $\mu$ and $T_c$ as
functions of $U$, the next step involves finding an efficient method 
to approximate 
$
\frac{\partial \mu}{\partial T}\big{|}_{T_c},
$
which is necessary for defining $\xi_\mathrm{GL}$ as it appears in 
Eq.~(\ref{eqn:a0}). From the usual BCS-BEC crossover problem, we know that
$
\frac{\partial \mu}{\partial T} \to 0
$
in the BCS limit, whereas it controls the dominant contribution to $a_0$ 
in the BEC limit~\cite{sademelo93}.
Thus, to properly capture the BEC limit, we recall the following insights 
from the Bogoliubov theory of a weakly-interacting (or low-density) atomic 
Bose gas~\cite{Fetter}: ($i$) the density of condensed bosons is given by
$
n_0(T) = n_\mathrm{B} [1 - (T/T_c)^\frac{3}{2}]
$
for $T < T_c$, and ($ii$) the chemical potential of superfluid bosons is 
given by
$
\mu_\mathrm{B}(T) = U_\mathrm{BB} n_0(T) > 0.
$
Here, $n_\mathrm{B}$ is the total density of condensed and noncondensed 
bosons, and $U_\mathrm{BB}$ is their onsite repulsion.
By making an analogy with $\mu_b = 2\mu - E_b$ of the Cooper pairs that 
is derived in Sec.~\ref{sec:sce}, we approximate
$
\frac{\partial \mu}{\partial T}\big{|}_{T_c} = 
\frac{1}{2} \frac{\partial \mu_p}{\partial T}\big{|}_{T_c}
= -3U_{pp} F_p/(4T_c),
$
where $U_{pp} = 2U$ is discussed below and $F_p = F_{bs}/2$. 

\begin{figure} [htb]
\includegraphics[width = 0.99\linewidth]{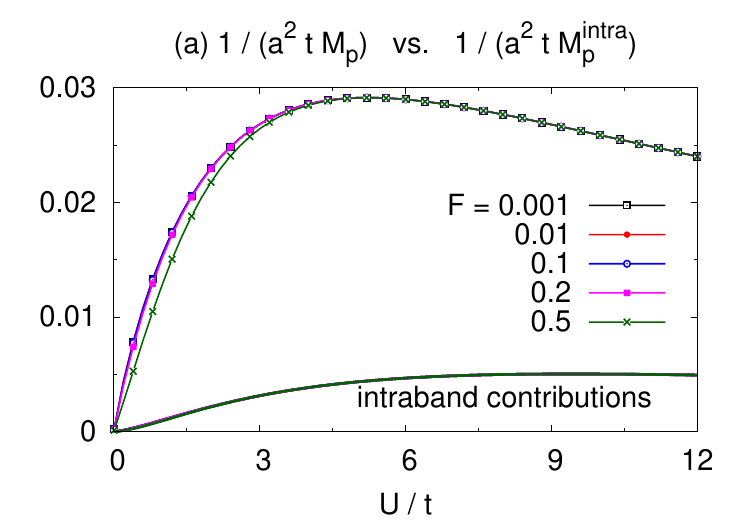}
\includegraphics[width = 0.99\linewidth]{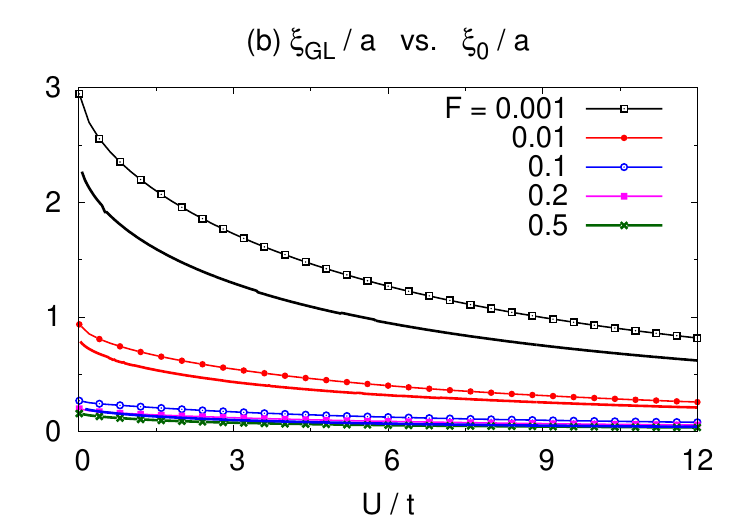}
\caption{\label{fig:TDGL}
The effective mass of the Cooper pairs (line plots with data points) and 
its conventional intraband contribution (line plots) are shown in (a) 
as functions of interaction strength for various fillings. Except for 
the $F = 0.5$ data, these are indistinguishable from the effective mass 
of the lowest-lying two-body bound states~\cite{iskin24} on the presented scale. 
Corresponding Ginzburg-Landau coherence lengths (line plots with data 
points) are shown in (b), and compared with the zero-temperature coherence 
lengths (line plots). The latter is not shown for $F = 0.2$ and $F = 0.5$ 
data where it is not valid. Typically, coherence length is inversely 
proportional with $U$ in the $U/t \gg 1$ regime.
}
\end{figure}

In Fig.~\ref{fig:TDGL}, we present self-consistent solutions for $1/M_p$ 
and $\xi_\mathrm{GL}$ as functions of $U/t$ for various particle 
fillings. Here, 
$
(M_p^{-1})_{ij} = \delta_{ij}/M_p
$ 
and
$
(\xi_\mathrm{GL}^2)_{ij} = \xi_\mathrm{GL}^2 \delta_{ij}
$
for the pyrochlore lattice as a consequence of uniform pairing.
Similar to the overall shape of the $T_c$ data, we observe that $1/M_p$ 
rises almost linearly for small $U \lesssim t$, it reaches a peak around 
$U \sim 5t$ and then decays as $t^2/U$ for $U \gg t$. This similarity
is reminiscent of the BEC transition temperature
$
T_c \propto n_\mathrm{B}^{2/3} / M_\mathrm{B}
$
of an ideal Bose gas~\cite{Fetter}.
Furthermore, Fig.~\ref{fig:TDGL}(a) shows that $1/M_p$ depends weakly 
on $F$ for all $U$, especially in the BEC limit when $F_{bs}/F \to 1$. 
This is because, by construction, our TDGL theory 
reproduces the exact effective mass $M_b$ of the lowest-lying two-body 
bound states ($i$) in the $F \ll 1$ limit for any $U/t \ne 0$, 
and ($ii$) in the $U/t \to \infty$ limit for any filling~\cite{iskin23}.
On the other hand, Fig.~\ref{fig:TDGL}(b) shows that $\xi_\mathrm{GL}$ 
decays to zero monotonously without bound as a function of $U/t$. 
There, we also present self-consistent solution for the zero-temperature 
coherence length $\xi_0$ as defined by Eq.~(\ref{eqn:xi0}), where
$
(\xi_0^2)_{ij} = \xi_0^2 \delta_{ij}
$
for the pyrochlore lattice due again to uniform pairing.
In the $F \ll 1$ limit, it is delightful to see that $\xi_\mathrm{GL}$ 
and $\xi_0$ length scales are approximately equal to each 
other, differing only by a prefactor of order unity. Thus, despite all of 
the approximations involved in the calculation of $\xi_\mathrm{GL}$, 
this benchmark demonstrates that our finite-temperature recipe produces 
accurate results for the dilute flat-band limit in the pyrochlore lattice.
See Appendix~\ref{sec:appA} for greater details about $\xi_0$ in the
pyrochlore lattice. In all of these observables, the conventional
intraband contributions play secondary roles for all $U$, 
as shown in Fig.~\ref{fig:TDGL}(a).

To gain further physical insight into these numerical observations, we 
examine two specific limits. For instance, in the $U/t \to 0$ limit when 
$\mu + 2t < 0$ and $|\mu + 2t| \gg T_c$, i.e., when $F \ll 1$, we find
$
\mu = -2t - U/4,
$
suggesting that $\xi_{n \mathbf{k}} \gg T_c$. Thus, by keeping only the 
flat-band contributions in the $\mathbf{k}$ sums, we find
$
a_0 = 3 F_{bs} / (8|\mu + 2t|)
$
for the static coefficient, where $F_{bs} \to F$ in this limit as shown 
in Fig.~\ref{fig:Tc}(a),
$
b_0 = 1/(8|\mu+2t|^3)
$
for the quartic coefficient,
$
c_{ij}^\mathrm{intra} = 0
$
and
$
c_{ij}^\mathrm{inter} = \frac{1}{2|\mu+2t| N} \sum_{f, m \notin f, \mathbf{k}}^{'}
g^{fm\mathbf{k}}_{ij}
$
for the kinetic coefficients, where $f = \{3,4\}$ refers to the flat bands 
and $m = \{1,2\}$ refers to the dispersive bands (the prime sum excludes 
the band touchings), and
$
d = 1/(8|\mu+2t|^2)
$
for the dynamic coefficient. They lead to
$
(M_p^{-1})_{ij} = \frac{|\mu+2t|}{N} \sum_{f, m \notin f, \mathbf{k}}^{'}
g^{fm\mathbf{k}}_{ij}
$
for the effective-mass of the Cooper pairs,
$
U_{pp} = 8|\mu+2t|
$
for the onsite repulsion between the pairs, and
$
(\xi_\mathrm{GL}^2)_{ij} = \frac{1}{6F_{bs} N} \sum_{f, m \notin f, \mathbf{k}}^{'}
g^{fm\mathbf{k}}_{ij}
$
for the GL coherence length among the pairs. It is pleasing to note that 
$U_{pp} \to 2U$ is consistent with the low-energy collective-mode analysis 
at $T = 0$~\cite{iskin24}. A more suggestive way to express the latter result is
$
\xi_\mathrm{GL} = \sqrt{2/(3 U_{pp} M_p F_p)},
$
which is approximately equal to the Bogoliubov expression
$
\xi_\mathrm{B} = \sqrt{1/(2 U_\mathrm{BB} M_\mathrm{B} F_\mathrm{B})}
$
of a weakly-interacting (or low-density) superfluid Bose gas~\cite{Fetter},
differing only by a prefactor of order unity. 
Similarly, in the same limit, we find
$
A = 4F(1-F)/U \to 4F/U
$
for the static coefficient, and recall $C_{ij} = c_{ij}/2$ as derived in
Sec.~\ref{sec:T0}. They lead to
$
\xi_0 = \sqrt{1/(4 U_{pp} M_p F_p)},
$
which again differs only by a prefactor of order unity. This result shows 
that $\xi_0 < \xi_\mathrm{GL}$, which is consistent with Fig.~\ref{fig:TDGL}(b).

In the other limit when $U/t \to \infty$, by noting that $\mu \to -U/2$ 
and
$
\xi_{n \mathbf{k}} \gg T_c,
$ 
we find
$
a_0 = 3F_{bs}/(4|\mu|)
$
for the static coefficient, where $F_{bs} \to F$ in this limit as shown
in Fig.~\ref{fig:Tc}(a),
$
b_0 = 1/(4|\mu|^3)
$
for the quartic coefficient,
$
c_{ij}^\mathrm{intra} = \frac{1}{4|\mu^3| N} \sum_{n \mathbf{k}} 
Q^{nn \mathbf{k}}_{ij}
$
for the conventional kinetic coefficient,
$
c_{ij}^\mathrm{inter} = \frac{1}{4|\mu^3| N} \sum_{n, m \ne n, \mathbf{k}} 
Q^{nm \mathbf{k}}_{ij}
$
for the geometric kinetic coefficient, and
$
d = 1/(4\mu^2)
$
for the dynamic coefficient. Here, we define
$
Q^{nm \mathbf{k}}_{ij} = \textrm{Re} [ 
\langle n_\mathbf{k}| \dot{\mathbf{h}}_\mathbf{k}^i |m_\mathbf{k} \rangle
\langle m_\mathbf{k}| \dot{\mathbf{h}}_\mathbf{k}^j | n_\mathbf{k} \rangle ],
$
where
$
\dot{\mathbf{h}}_\mathbf{k}^i = \partial \mathbf{h}_\mathbf{k} / \partial k_i
$
is the derivative of the Bloch Hamiltonian. This leads to
$
Q^{nn \mathbf{k}}_{ij} = \dot{\varepsilon}_{n\mathbf{k}}^i \dot{\varepsilon}_{n\mathbf{k}}^j
$
for the intraband processes, and
$
Q^{n \ne m \mathbf{k}}_{ij} = \frac{1}{2}(\varepsilon_{n\mathbf{k}} - \varepsilon_{m\mathbf{k}})^2 g^{nm\mathbf{k}}_{ij}
$
for the interband processes.
In deriving $c_{ij}^\mathrm{inter}$, we also observed that
$
\sum_{n, m \ne n, \mathbf{k}}
(\varepsilon_{n\mathbf{k}} - \varepsilon_{m\mathbf{k}}) 
g^{nm\mathbf{k}}_{ij} = 0
$
and
$
\sum_{n, m \ne n, \mathbf{k}}
(\varepsilon_{n\mathbf{k}}^2 - \varepsilon_{m\mathbf{k}}^2) 
g^{nm\mathbf{k}}_{ij} = 0
$
since
$
g^{nm\mathbf{k}}_{ij} = g^{mn\mathbf{k}}_{ji}
$
and
$
g^{n\mathbf{k}}_{ij} = g^{n\mathbf{k}}_{ji}
$
by definition. Thus, the total kinetic coefficient can be calculated 
via the identity
$
\sum_{nm \mathbf{k}}
\langle n_\mathbf{k}| \dot{\mathbf{h}}_\mathbf{k}^i | m_\mathbf{k} \rangle
\langle m_\mathbf{k}| \dot{\mathbf{h}}_\mathbf{k}^j | n_\mathbf{k} \rangle
= \sum_\mathbf{k} \mathrm{Tr}[\dot{\mathbf{h}}_\mathbf{k}^i 
\dot{\mathbf{h}}_\mathbf{k}^j]
= N_c a^2 t^2 \delta_{ij},
$
suggesting that
$
c_{ij} = a^2 t^2 \delta_{ij}/(16|\mu|^3).
$
Together with the rest of the coefficients, they lead to
$
(M_p^{-1})_{ij} = a^2 t^2 \delta_{ij}/(4|\mu|)
$
for the effective-mass of the Cooper pairs,
$
U_{pp} = 4|\mu|
$
for the onsite repulsion between the pairs, and
$
(\xi_\mathrm{GL}^2)_{ij} = a^2 t^2 \delta_{ij}/(12 U |\mu| F)
$
for the GL coherence length among the pairs. It is pleasing to note that 
$M_p = 2U/(a^2 t^2)$ and $U_{pp} \to 2U$ are again consistent with the 
low-energy collective-mode analysis at $T = 0$~\cite{iskin24}. 
The latter result can again be written in a more suggestive way
$
\xi_\mathrm{GL} = \sqrt{1/(3 U_{pp} M_p F_p)}.
$
Similarly, in the same limit, we find
$
A = F(2-F)/U \to 2F/U
$
for the static coefficient, and recall that $C_{ij} = c_{ij}/2$, leading to
$
\xi_0 = \sqrt{1/(4 U_{pp} M_p F_p)}.
$
Thus, we again conclude that $\xi_\mathrm{GL}$ and $\xi_0$ differ from 
each other and from $\xi_\mathrm{B}$ by factors of order unity, reflecting 
the dilute-limit behavior. Moreover, $\xi_0 < \xi_\mathrm{GL}$, 
which is consistent with Fig.~\ref{fig:TDGL}(b).

\section{Discussion}
\label{sec:disc}

Our self-consistent GL analysis near $T_c$ and the fluctuation analysis 
at zero temperature both indicate that the coherence length is not 
limited by a fundamental length scale determined by the quantum geometry 
of the flat band, which contrasts sharply with the recent proposal by 
Hu et al.~\cite{hu23}. We believe that the primary source of this 
discrepancy lies in their projection scheme, specifically the projection 
of the fermionic operators onto the flat band. Under the assumptions 
of time-reversal symmetry and uniform pairing, our TDGL coefficients 
near $T_c$ that are given by Eqs.~(\ref{eqn:a0})-(\ref{eqn:d}) and
our fluctuation analysis at zero temperature that are given by 
Eqs.~(\ref{eqn:A})-(\ref{eqn:Cinter}) are both applicable to any 
tight-binding lattice without constraints on the resultant band structure.
In particular, our results are applicable to all $U \ne 0$.
On the other hand, Hu et al. focus on systems with an isolated flat band 
that is energetically separated from other bands by energy gaps both 
above and below. Their projection of the mean-field Hamiltonian onto 
the subspace of the so-called flat-band fermions restricts their results 
to the low $U/t$ regime. Our results, however, demonstrate that other 
bands play equally significant roles in the $U/t \gg 1$ regime, 
which aligns with intuitive expectations. As a result, we conclude 
that correlation lengths, such as the coherence length, can indeed be 
smaller than one lattice spacing without being constrained by the 
quantum metric. This conclusion is consistent with the more recent 
findings of Ref.~\cite{thumin24}.

\section{Conclusion}
\label{sec:conc}

In summary, we developed a self-consistent formulation of the GL coherence 
length $\xi_\mathrm{GL}$ in a dilute flat-band superconductor using 
the multiband pyrochlore-Hubbard model. Our results are highly consistent 
with the zero-temperature coherence length $\xi_0$, demonstrating that 
$\xi_\mathrm{GL}$ decreases to zero monotonously as $U/t$ increases. 
Furthermore, we established that the effective mass of Cooper pairs 
aligns closely with that of the lowest-lying two-body bound states in 
the dilute regime. As an outlook, we plan to improve the quantitative 
accuracy of these findings by implementing a more accurate number 
equation near $T_c$. For instance, similar to the crucial role played 
by the bound-state contribution $F_{bs}$ in the dilute flat-band limit, 
we expect that the scattering-state contribution $F_{sc}$ will become 
equally important at and around the half-filled flat-band limit. Although 
we do not expect any qualitative change, inclusion of $F_{sc}$ is inevitable 
there, especially in the low-$U/t$ regime. Furthermore, there is room
for improvement in the calculation of $F_{bs}$, e.g., by extracting 
$\omega_{b \mathbf{q}}$ more accurately without relying on the two-body 
results, or by incorporating contribution from other poles.

\begin{acknowledgments}
The author acknowledges funding from US Air Force Office of Scientific 
Research (AFOSR) Grant No. FA8655-24-1-7391.
\end{acknowledgments}

\appendix

\section{Breakdown of $(\xi_0^2)_{ij}$} 
\label{sec:appA}

\begin{figure} [htb]
\includegraphics[width = 0.99\linewidth]{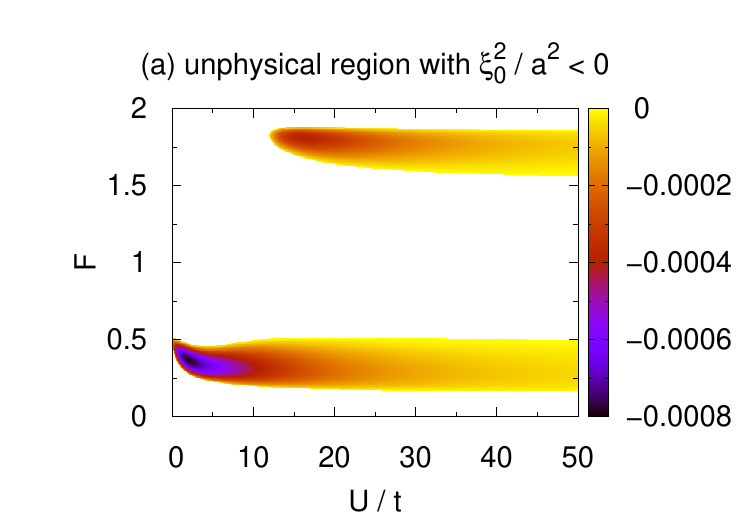}
\includegraphics[width = 0.99\linewidth]{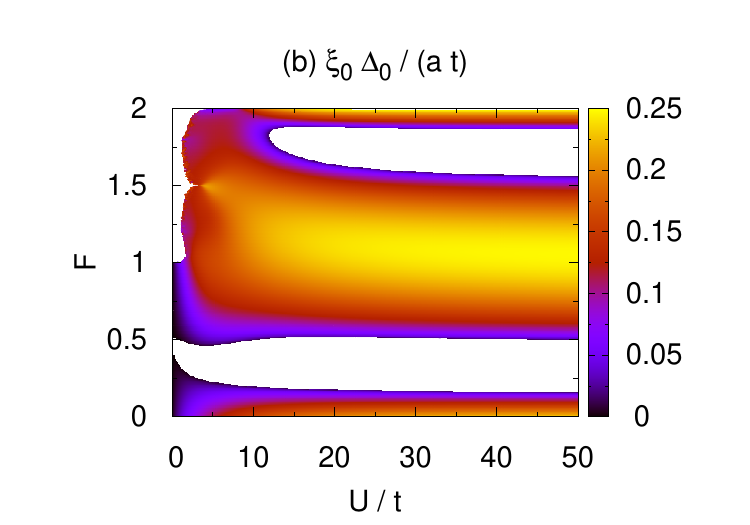}
\caption{\label{fig:app}
The colored map of zero-temperature coherence length is shown in (a)
for the unphysical region where $\xi_0^2 < 0$. 
In (b), $\xi_0/a$ is shown to scale as $t/\Delta_0$ in most of the physical 
region, except for the flat-band superconductivity in the low-$U/t$ regime. 
Since our numerics becomes unreliable in the $\Delta_0/t \to 0$ limit, 
we present only the data with $\Delta_0/t > 0.001$, revealing 
the underlying single-particle density of states at the periphery of the 
white region~\cite{iskin24} in (b), i.e., for the BCS limit when $1 < F <2$. 
} 
\end{figure}

By definition, the zero-temperature coherence length $(\xi_0^2)_{ij}$ is 
not well-defined in the parameter regime where $C_{ij}$ is negative. 
In Fig.~\ref{fig:app}(a), we present $\xi_0^2 < 0$ values in the unphysical 
region as a function of $U/t$ and $F$, indicating that $\xi_0$ can be 
meaningfully defined for the dilute flat-band ($F \ll 1$) limit of our 
interest in the pyrochlore lattice. Similar to the 
well-studied continuum model with a quadratic dispersion~\cite{pistolesi96},
this is known to be the case for a dilute single-band lattice model with 
cosine dispersion~\cite{benfatto02}, where $\xi_0$ and $\xi_\mathrm{GL}$ 
are known to give very similar results up to a constant of order unity for 
all $U \ne 0$. When $C_{ij} < 0$, it signals that the minimum of 
$
\mathcal{M}^\mathbf{q}_{11} + \mathcal{M}^\mathbf{q}_{12}
$
occurs at a finite $\mathbf{q}$. If desired, one can perform the 
low-$\mathbf{q}$ expansion around the new minimum instead of 
$\mathbf{q = 0}$, and define $(\xi_0^2)_{ij}$ accordingly.

On the other hand, in Fig.~\ref{fig:app}(b), we present $\xi_0$ in the
physical region as a function of $U/t$ and $F$, showing that $\xi_0/a$ 
scales with $t/\Delta_0$ in most of the parameter space, except for the 
flat-band superconductivity in the low-$U/t$ regime. There, $\xi_0$ is
governed solely by the quantum geometry of the Bloch states. We note that a similar 
scaling has recently been reported in the arxiv for a distinct but a related
correlation length in the context of the sawtooth lattice~\cite{thumin24}.
Thus, similar to the dilute case, $\xi_0$ decreases without bound as 
$U/t$ increases for any $F$ within the physical region.

\bibliography{refs}

\end{document}